# Strain-assisted optomechanical coupling of polariton condensate spin to a micromechanical resonator

O. Beer,[1] H. Ohadi,[1*] Y. del Valle-Inclan Redondo,[1] A. J. Ramsay,[2] S. I. Tsintzos,[3] Z. Hatzopoulos,[3] P. G. Savvidis,[3,4,5] J. J. Baumberg[1†]

[1] *NanoPhotonics Centre, Department of Physics, Cavendish Laboratory, University of Cambridge, Cambridge CB3 0HE, United Kingdom*

[2] *Hitachi Cambridge Laboratory, Hitachi Europe Ltd., Cambridge CB3 0HE, United Kingdom*

[3] *FORTH, Institute of Electronic Structure and Laser, 71110 Heraklion, Crete, Greece*

[4] *Department of Materials Science and Technology, University of Crete, 71003 Heraklion, Crete, Greece*

[5] *ITMO University, St. Petersburg 197101, Russia*

We report spin and intensity coupling of an exciton-polariton condensate to the mechanical vibrations of a circular membrane microcavity. We optically drive the microcavity resonator at the lowest mechanical resonance frequency while creating an optically-trapped spin-polarized polariton condensate in different locations on the microcavity, and observe spin and intensity oscillations of the condensate at the vibration frequency of the resonator. Spin oscillations are induced by vibrational strain driving, whilst the modulation of the optical trap due to the displacement of the membrane causes intensity oscillations in the condensate emission. Our results demonstrate spin-phonon coupling in a macroscopically coherent condensate.

## I. INTRODUCTION

There is a growing interest in coupling vibrations to other degrees of freedom in order to probe, sense, and cool mechanical resonators. Examples include coupling of cavity phonons to cavity photons[1], effects of vibration on photocurrent[2], vibration-based logic gates[3], coupling of vibrations to the spin of single nitrogen–vacancy centers[4], magnetization reversal in molecular magnets[5], ferromagnetic resonance[6] or spin transport[7] by surface acoustic waves, and vibrational coupling of quantum dots to photonic crystal cavities[8]. An emerging opportunity is to vibration-modulate polaritons - these are quasiparticles that result from the strong coupling of cavity photons to quantum-well (QW) excitons[9], and can condense into macroscopic quantum states[10–12]. Mechanical effects on microcavity exciton-polaritons have been demonstrated before, either by applying static pressure[13], or inducing acoustic waves over the microcavity surface[14]. Recently, optomechanical feedback coupling through the nonlinearity of the polariton system has also been proposed[15]. However, the coupling of the spin (or magnetic moment) of a condensate to vibrations has not yet been achieved.

Here we demonstrate coupling of the occupation (intensity) and spin (polarization) of an optically trapped polariton condensate to the lowest vibrational mode of a circular membrane microcavity (forming an acoustic resonator) [FIG.1(a)]. We find that polarization modulation is mediated by strain, whereas condensate intensity varies through the modulation of the optical trap due to displacement of the membrane. Our work is a demonstration of the coupling of spin to vibrations in a quantum many-body

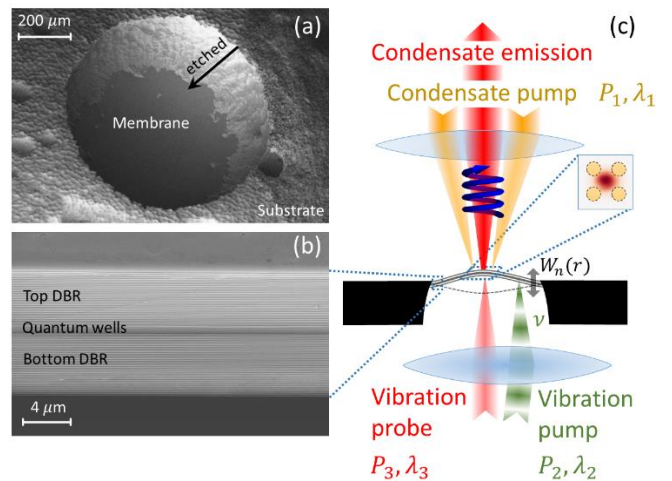

FIG. 1. (a) Electron microscope image of the microcavity membrane, taken from the etched backside. (b) Cross section of the membrane. (c) Experimental setup. Above the membrane (front side), orange beams form the four-spot polariton trap giving circularly polarized polariton emission (red). Below the membrane (backside), green beam is the amplitude modulated vibration-inducing pump, pink beam is the vibration probe.

system and has potential applications in strain sensing [8,16] and spin-chain interactions[17].

We create a trapped polariton condensate by non-resonantly optical pumping a GaAs QW microcavity[18,19]. The polariton pump ($P_1 = 20$ mW, $\lambda_1 = 765$ nm) is patterned into a 4-spot trap using a spatial light modulator (SLM) [FIG.1(b)]. This non-resonant polariton pump is blue detuned from the optical resonance of the microcavity creating a reservoir of hot excitons at each pump spot. The reservoir excitons spread out due to repulsive Coulomb interactions and rapidly lose energy until they form polaritons. The polariton pump provides both gain medium and trapping potential[20,21]. Above a critical threshold polaritons spontaneously condense into a macroscopically coherent state inside the trap due to stimulated scattering[22].

Polariton spin is quantized ($\pm 1$) along the growth axis of the microcavity, emitting photons with left- ($\sigma^-$) or right ($\sigma^+$) circular polarization when they decay. Above a spin-bifurcation power threshold (under appropriate conditions), condensed polaritons spontaneously magnetize into specific randomly-selected spin states. The spin-bifurcation process is driven by the dissipation rate difference ($\gamma$) and energy splitting ($\varepsilon$) of the horizontally and vertically polarized polaritons, or birefringence[19]. We operate above this threshold, where the degree of circular polarization (condensate spin) $s_z$ exceeds 90% (FIG. 2). By slightly changing the polarization of the polariton pump laser to left- (or right-) circularly polarized (<5%), we can deterministically form spin-down (or up) polariton condensates.

We measure the circular degree of polarization $s_z = (I_{\sigma^+} - I_{\sigma^-})/(I_{\sigma^+} + I_{\sigma^-})$ and the total intensity $I = (I_{\sigma^+} + I_{\sigma^-})$ of the condensate in the membrane microcavity while optically exciting the fundamental vibrational mode of the resonator. The condensate spin, which is the z-component of the polarization vector, varies quadratically with the energy splitting, $s_z \propto \varepsilon^2$, which is now modulated by the periodically-driven strain[13,19].

## II. EXPERIMENTAL SETUP

The microcavity is composed of two distributed Bragg reflector (DBR) mirrors (32 layers) which form a 5/2λ cavity sandwiching 10-nm GaAs QWs placed at the maxima of the cavity light field (see[18] for more details). To form a mechanical resonator, the 300 μm thick GaAs substrate of the microcavity is chemically etched to release ~700 μm-diameter membranes of thickness $h$=8.7 μm [FIG. 1(a,b)].

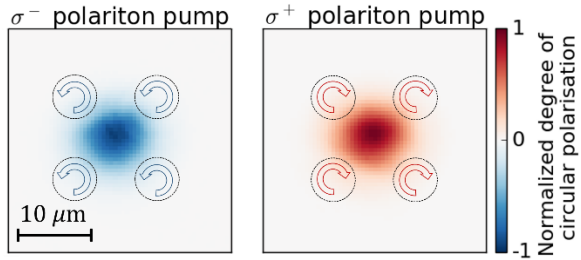

FIG. 2. Experimental emission from the two possible states of the trapped polariton condensate: spin-up (right-circular polarized, red) or spin-down (left-circular polarized, blue). Circles show slightly circular polarized polariton pump spots.

Membrane microcavities are mounted inside a cryostat with optical access from both sides of the sample. To mechanically excite the resonator, the backside of the sample is periodically heated with an amplitude-modulated continuous wave (CW) Ti:Sapphire laser ('vibration-pump' $P_2 \simeq 4$ mW [FIG.1(c)]. To maximise absorption in the resonator, the vibration pump laser wavelength $\lambda_2$=765-775 nm is blue detuned from the microcavity stop-band[23]. The acousto-optic modulation frequency is then tuned through the vibration resonance.

For vibration detection, we use a diode laser ('vibration-probe' $P_3 \simeq 1$ mW) from the back side. The vibration probe $\lambda_3$=800 nm is set to the microcavity stop-band reflection maximum. The reflected probe beam is interfered with a reference beam on the detection photodiode. By calibrating the maximum amplitude of interference signals, the membrane displacement can be accurately measured. The vibration pump is located more than 200 μm from the polariton condensate to avoid any additional polariton injection.

We first create condensates with fixed circular polarization (that persists while the pump is on) by tuning the circular polarization of the polariton pump (FIG. 2). The circular degree of condensate polarization is measured using a polarization beamsplitter and two photomultipliers. All three lasers (polariton pump, vibration-pump, and vibration-probe) are turned on for each 130 ms measurement. The total intensity ($I$) and the circular polarization ($s_z$) of the condensate emission are measured at different positions across the microcavity membrane as the frequency of the vibration pump is scanned.

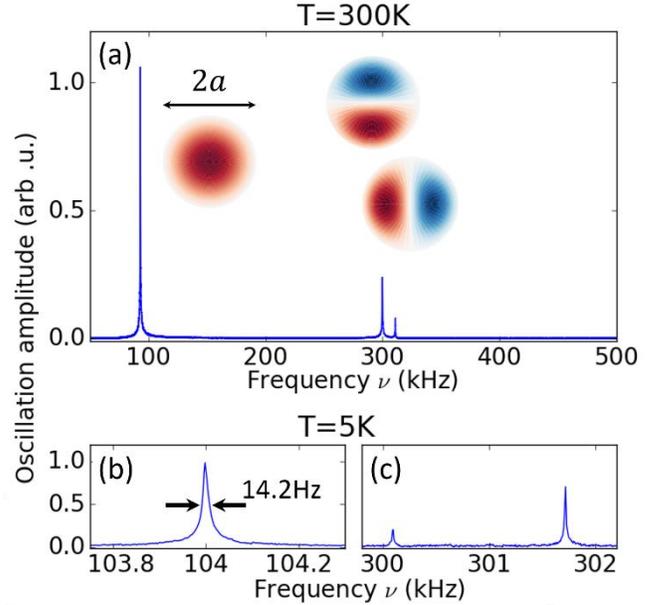

FIG. 3. Mechanical vibration spectrum of the microcavity membrane with the first three modes and their simulated displacements at (a) room temperature, and (b,c) at cryogenic temperatures acquired using lock-in detection.

## III. RESULTS AND DISCUSSION

To find the resonance frequencies of the membrane we first perform a frequency scan at room temperature [FIG. 3(a)]. A resonance is observed by the vibration-probe whenever the modulation frequency of the vibration-pump laser coincides with a mechanical resonance of the membrane. Vibration is induced by the photothermal effect[24]. The highest vibration frequency that can be excited in the microcavity is $f_{max} = 1/\tau = 4\pi D/r^2 \simeq 1.5$ GHz, where $\tau$ is the relaxation time, $D = 3.1 \times 10^{-3}$ m$^2$/s is the thermal diffusivity of GaAs, and $r = 5 \times 10^{-6}$ m is the radius of the vibration-pump spot. While the mechanical quality-factor of the resonator at room temperature is low (~300), at cryogenic temperatures (T = 5K) the quality factor of the fundamental mode at $f_0$=104 kHz rises to ~7300 [FIG. 3(b, c)]. At this resonance a vibration amplitude of 3 nm is measured at the center of membrane using our calibrated interferometer [FIG. 4(b)].

The membrane displacement and strain are treated as those of a clamped plate. The $n^{th}$ mode spatial displacement at membrane position $(r, \theta)$ due to free vibration is then[25]:

$$W_n(r,\theta) = C_n \left[ -\frac{I_n(k_n a)}{J_n(k_n a)} J_n(k_n r) + I_n(k_n r) \right] \cos(n\theta) \quad (1)$$

with $k_n$ defined by

$$k_n^2 = \omega_n \sqrt{\rho h/D}. \quad (2)$$

Here $C_n$ is proportional to the maximal mechanical oscillation amplitude, $n$ is the mode number, $J_n$ ($I_n$) is the (modified) Bessel function, $a$=350 µm is the plate radius, $\rho$ is the mass density of the membrane (4450 kg/m$^3$), $D = Eh^3/[12(1 - \nu^2)]$ is the flexural rigidity, with Young's modulus of the (100) plane along the [110] direction $E = 80$ GPa[26], and with Poisson ratio $\nu = 0.3$. Using these values, gives $f_0 = \omega_0/2\pi = 136$ kHz (and $k_0 a$=3.2) which is larger than the measured 104 kHz. This discrepancy suggests that there is some radial tension in the membrane. The calculated displacement patterns of the first two modes ($W_{0,1}$) are plotted in FIG. 3(a). The radial strain for the fundamental mode [FIG. 4(a)] is

$$\epsilon_{rr} = \frac{1}{2}\left(\frac{\partial W_0}{\partial r}\right)^2. \quad (3)$$

As shown in FIG. 4(a), this strain is zero at the center of the membrane and maximized at $r = a/2$.

The mechanical resonance is observed in measurements of the condensate circular polarization and total intensity [FIG. 4(b)]. This coupling of mechanical vibration to condensate polarization and intensity is position dependent and governed by strain and displacement. To extract their amplitudes, the Fourier transforms of $I, s_z$ are taken as the driving frequency is scanned [FIG. 4(c)]. Sidebands in intensity are due to intensity modulations of the SLM (~1 kHz). When the polariton condensate is located 100 µm from the center of the membrane at $r \sim a/2$ [FIG. 4(a), left] the displacement modulation is small, whereas the strain modulation is maximal. For this reason, the polarization-modulation is strongest while the intensity modulation is reduced. On the other hand, at the center of the membrane ($r$=0) where the displacement is maximum and strain is minimum [FIG. 4(a), right] the polarization oscillation is also minimum and the intensity oscillation is maximum. In the optimal off-center position ($r \sim a/2$) a maximum polarization modulation of ~1.2% is observed where the radial strain amplitude is maximum ($\max(\epsilon_{rr}) = 4 \times 10^{-11}$). We now explain how this can yield the modulation of the condensate embedded in the vibration multilayer.

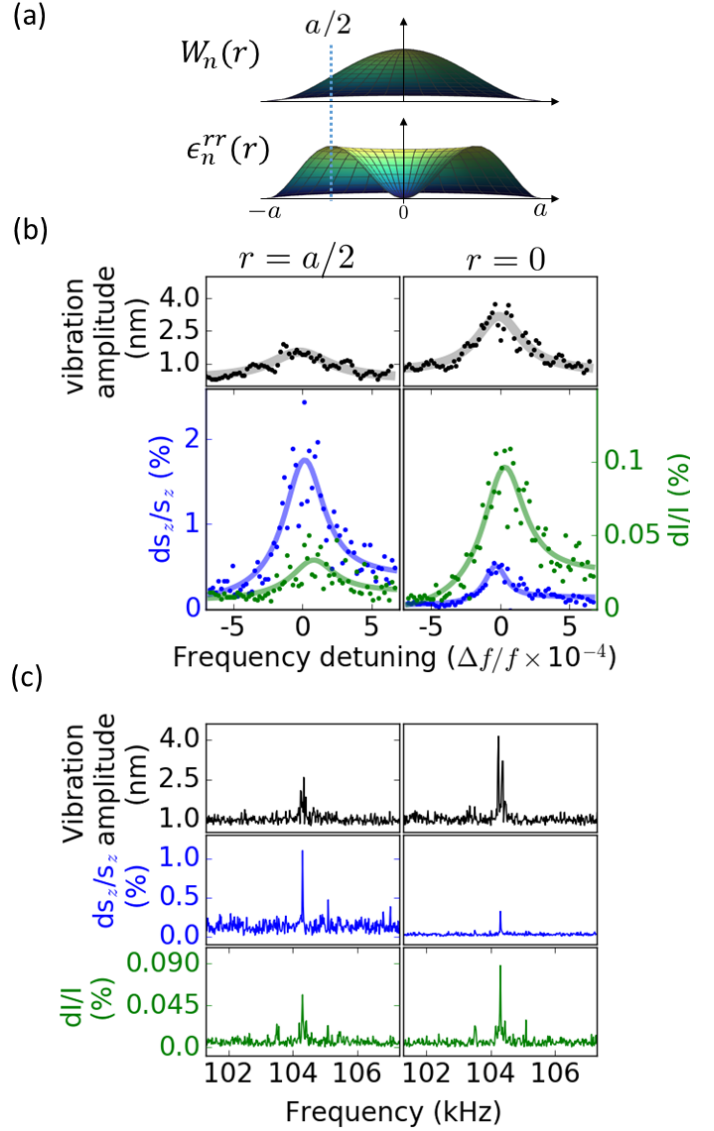

FIG. 4. (a) Displacement ($W_n$) and radial strain $\epsilon_n^{rr}$ of the fundamental vibration mode of the membrane. (b) Amplitude of Fourier component at mechanical drive frequency measured at two positions $r = 0, a/2$. (top) Membrane displacement (bottom, blue) Polarization of condensate (bottom, green) Intensity of condensate. Gray lines are Lorentzian fits to the mechanical oscillation and blue and green lines are moving averages of polarization and intensity. Each dot is the maximum of the power spectrum at the vibration frequency. (c) Power spectrum of the membrane vibration and polariton polarization and intensity around the resonance of mechanical driving, extracted from the Fourier transform of the corresponding time traces.

The modulation of spin due to membrane displacement is described using a zero-dimensional (0D) model of polariton condensates which includes the spin bifurcation (see [19] for details). In this case, the birefringence is modulated by the mechanical vibration at a frequency $f$:

$$\varepsilon = \varepsilon + \delta\varepsilon \cos(2\pi f_0 t), \quad (4)$$

where $\delta\varepsilon$ is the modulation depth. For small $\delta\varepsilon$, the spin modulation depth $\delta s_z/s_z$ is proportional to polarization energy splitting modulation depth $\delta\varepsilon/\varepsilon$. This gives an estimated birefringence modulation ($\delta\varepsilon/\varepsilon$) of ~1% in the microcavity resonator. The sensitivity of the strain in our experiment ($S_\epsilon$) can be calculated from the measurement bandwidth ($BW$~10 Hz), and the maximum applied strain modulation [$\max(\epsilon_{rr}) = 4 \times 10^{-11}$] by $S_\epsilon = \frac{\max(\epsilon_{rr})}{\sqrt{BW}} \simeq 1.3 \times 10^{-11} \, 1/\sqrt{\text{Hz}}$. This confirms that opto-magneto-mechanical coupling of condensates with membrane vibrations can be achieved. It thus provides a way to control spin evolution in these condensates by appropriately engineering the strain distribution and dynamics.

The intensity modulation of the condensate when it is located around the center of the membrane is due to changes in the optical trapping. The change in the separation between two adjacent pump spots due to a 3nm displacement of the membrane is estimated by geometrical optics to be ~50 pm, while the overall change in the diameter of the Gaussian spots is negligible at the focus of the objective (<0.01pm). To estimate how the change in the trap size affects the condensate intensity we performed 2-dimensional Ginzburg-Landau simulations[27]. This shows that a 30 pm change in the 13 $\mu$m pump spot separation used here, is sufficient to cause the 0.1% modulation in condensate intensity observed in the experiment. The stimulated scattering of polaritons into the condensate, and hence the intensity of the condensate is highly sensitive to small changes in trap size (order $10^{-6}$). Hence, a polariton condensate may have potential as an ultrasensitive microphonic sensor.

## IV. CONCLUSIONS

We demonstrate position-dependent intensity and spin modulation of an optically trapped polariton condensate due to mechanical vibrations of a circular membrane microcavity resonator. The intensity is directly modulated by the displacement of the membrane, which for the lowest mode is maximum at the center $r=0$. Displacement of the membrane modulates the depth of the optical trap, which in turn modulates the occupation and intensity of the condensate. The spin is modulated by the strain and is maximum off-center (at $r=a/2$). The strain modulates the birefringence of the cavity $\varepsilon$, and can be described by the spin-bifurcation model. Our experiment is sensitive to few-nanometer displacements and <0.1% changes in strain and could potentially be used as a sensor. Furthermore, by using smaller microcavity resonators, strong coupling of vibrations and spin may be achievable.

## ACKNOWLEDGEMENTS

We acknowledge Grants No. EPSRC EP/L027151/1, ERC LINASS 320503, and Leverhulme Trust Grant No. VP1-2013-011. PS acknowledges support from ITMO Fellowship Program and megaGrant No. 14.Y26.31.0015 of the Ministry of Education and Science of Russian Federation. AJR acknowledges support of Horizon 2020 programme (No. FETPROACT-2016 732894-HOT). Supporting research data can be found at: doi.org/10.17863/CAM.16887

* ho278@cam.ac.uk
† jjb12@cam.ac.uk